\documentclass[prx, twocolumn, 10pt, aps, floatfix, nofootinbib, nobibnotes, superscriptaddress, longbibliography, groupedaddress]{revtex4-2}
\usepackage{siunitx}
\usepackage[utf8]{inputenc}
\usepackage{orcidlink}
\usepackage{amsmath,amsfonts,amssymb}
\usepackage{amsmath}
\usepackage{color}
\usepackage{soul} 
\usepackage{subfigure}
\usepackage{multirow}
\usepackage{dcolumn} 
\usepackage{bm} 
\usepackage{graphicx} 
\usepackage{hyperref} 
\hypersetup{
    colorlinks={true},
    linkcolor={blue},
    citecolor={blue},
    urlcolor={blue}
}

\usepackage{xcolor}         
\usepackage[normalem]{ulem}
\usepackage{pifont}
\usepackage{makecell}

\newcommand{\pd}{\partial}
\newcommand{\lt}{\left}
\newcommand{\rt}{\right}

\newcommand{\nn}{\nonumber}

\newcommand{\bs}{\boldsymbol}

\newcommand{\mc}{\mathcal}
\newcommand{\mbf}{\mathbf}

\newcommand{\ra}{\rangle}
\newcommand{\la}{\langle}

\newcommand{\td}{\widetilde}

\newcommand{\ve}{\varepsilon}
\newcommand{\tve}{\widetilde{\varepsilon}}

\newcommand{\bk}{\boldsymbol{k}}

\newcommand{\et}{\boldsymbol{E}^T}

\newcommand{\tbc}{\boldsymbol{\mathcal{\tilde{A}}}}

\newcommand{\tbcp}{\mathcal{F}}

\newcommand{\og}{\bs{\Omega}}

\begin{document}

\title{Intrinsic Nonlinear Planar Thermal Hall Effect}

\author{Chanchal K. Barman}
\email{arckb2@gmail.com}
\affiliation{Dipartimento di Fisica, Universit\`a di Cagliari, Cittadella Universitaria, Monserrato (CA) 09042, Italy}

\begin{abstract}
We introduce the \textit{intrinsic} nonlinear planar thermal Hall effect (NPTHE)-- a dissipationless thermal response proportional to $(\nabla T)^2B$, which arises when the temperature gradient $\bs{\nabla} T$ and magnetic field $\mathbf{B}$ lie within the same plane.
The effect originates from a thermal gradient induced correction to the Berry curvature, characterized by the thermal Berry connection polarizability (TBCP) tensor, leading to a nonlinear transverse heat current independent of scattering time.
A symmetry analysis shows that the \textit{intrinsic} NPTHE is permitted only in noncentrosymmetric crystal point groups lacking horizontal mirror symmetry.
Using a tilted Dirac model, we demonstrate that its characteristic angular dependence provides an effective means to control the nonlinear thermal response.
Our results establish a new class of quantum geometry driven \textit{intrinsic} nonlinear thermal transport, offering both a sensitive probe of band geometry and a pathway toward nonlinear thermal functionalities in quantum materials.
\end{abstract}

\maketitle

\noindent \textcolor{black}{\textit{Introduction}}--- Transport phenomena governed by the quantum geometric properties of Bloch bands have emerged as a central theme in modern condensed matter physics. Over the past decade, concepts such as Berry curvature, orbital moment, and quantum metric have been recognized as key ingredients determining a wide class of Hall-type responses, even in systems without external magnetic fields. These geometrical quantities encapsulate the topological and symmetry characteristics of the underlying crystal, giving rise to effects such as the anomalous Hall effect, anomalous Nernst effect, and nonlinear Hall transport in both magnetic and nonmagnetic materials \cite{Xiao2010Berry,Nonlinear2021Du,Nagaosa2010Anomalous}. These discoveries have established a unified language connecting topology, symmetry, and dissipationless response. These responses arise from two distinct origins: \textit{intrinsic} mechanisms, rooted in the band geometry itself, and \textit{extrinsic} mechanisms, which rely on scattering-induced asymmetries of charge carriers \cite{sinitsyn2007semiclassical,Nonlinear2021Du,Nagaosa2010Anomalous,Quantum2021Du}. Particularly, the well-known linear anomalous Hall effect is a purely \textit{intrinsic} phenomenon governed solely by the Berry curvature---a fundamental band geometric quantity---and is independent of the scattering time $\tau$ \cite{Nagaosa2010Anomalous,Jungwirth2002AHE}. In contrast, the second-order anomalous Hall effect, linked to the Berry curvature dipole, explicitly depends on scattering time $\tau$ and is therefore regarded as an \textit{extrinsic} contribution \cite{Sodemann2015BCD}.

Recently, nonlinear Hall phenomena have attracted considerable attention as they extend the concept of Berry-curvature–driven transport beyond the linear-response regime. In particular, the discoveries of the \textit{intrinsic} second-order anomalous Hall effect \cite{liu2021Intrinsic,wang2021Intrinsic}, the \textit{intrinsic} second-order valley Hall effect \cite{kamaldas2024valleyHall}, and the \textit{extrinsic} third-order anomalous Hall effect \cite{lai2021third,tanay2023thirdorder,barman2025thirdorder,BCP2022Liu} have revealed that the Berry curvature itself can be field-corrected by an applied electric field through the so called Berry connection polarizability (BCP) tensor or the renormalized quantum metric \cite{Gao2014Field,yu2025quantum,gao2023quantum}. Such field-induced corrections expand the established taxonomy of \textit{intrinsic} Hall responses beyond the linear order. Extending this framework, Huang et al.\,\cite{IPHE2023} identified an \textit{intrinsic} nonlinear planar Hall effect, originating from the BCP.
In this effect, the Hall current scales as $j_H \propto E^2 B$, with $E$ representing the driving electric field and $B$ an in-plane magnetic field that couples to the band geometry.

Thermal transport, governed by analogous band geometric principles, offers a natural arena for extending these concepts. It is well established that the anomalous thermal Hall and Nernst effects originate from the Berry curvature weighted by the energy distribution of carriers. However, their nonlinear thermal counterparts remain largely less explored, especially those arising intrinsically from corrections to the quantum geometric structure of Bloch states rather than from \textit{extrinsic} scattering processes.
Analogous to the electric field induced correction of Berry curvature, it has recently been shown that a temperature gradient can also generate a correction to the Berry curvature, characterized by a corresponding thermal Berry connection polarizability (TBCP) \cite{li2024TBCP,zhang2025TBCP}.

Motivated by this insight, in this work we propose and develop the concept of an \textit{intrinsic} nonlinear planar thermal Hall effect (NPTHE) -- a nonlinear, dissipationless heat current that arises transverse to a temperature gradient as a consequence of a thermal gradient induced correction to the Berry curvature. The planar thermal Hall effect (PTHE) refers to a configuration in which the applied magnetic field ($B$) and temperature gradient ($\nabla T$) both lie within the transport plane that also hosts the transverse heat current $j^Q_H$.
Unlike the conventional thermal Hall effect, where an out-of-plane magnetic field drives a Lorentz-type deflection of carriers, the PTHE occurs entirely under in-plane conditions, where the Lorentz force plays no role. Instead, the \textit{intrinsic} NPTHE solely originates from the interplay between the in-plane magnetic field and the band geometry, manifesting through a thermal gradient induced correction to the Berry curvature mediated by the thermal Berry connection polarizability (TBCP) \cite{li2024TBCP,zhang2025TBCP} and its spin susceptibility \cite{IPHE2023}. This planar configuration provides a clean platform to probe \textit{intrinsic} quantum geometric contributions to nonlinear thermal Hall transport, independent of \textit{extrinsic} scattering effects.

\noindent \textcolor{black}{\textit{Theoretical formalism and origin of intrinsic NPTHE}}--- To uncover the microscopic origin of the \textit{intrinsic} NPTHE, we consider a nonmagnetic system subjected to an in-plane temperature gradient $\nabla T$ and a magnetic field $B$ in a coplanar $x$-$y$ configuration. Since the heat current is odd under time reversal $\mc{T}$, while $\nabla T$ is even, the \textit{intrinsic} transport current cannot arise without the presence of the magnetic field. In this coplanar configuration, the magnetic field couples to the electron’s magnetic moment through a \textit{Zeeman}-like interaction. Consequently, the effect of the magnetic field can be treated as a perturbation to the band structure, inducing spin splitting and effectively converting the original nonmagnetic bands into a ``magnetic'' band structure. The local Hamiltonian incorporating the \textit{Zeeman} term can be written as $\hat{H}_c = \hat{H}_0 + \frac{g}{\hbar}\bs{\mu}_B\cdot \mbf{B}$, where $\hat{H}_0$ is the unperturbed Hamiltonian, and $g$ and $\mu_B$ are the Land\'e $g$-factor and Bohr magneton respectively. The magnetic field couples to the electron spin, thereby modifying both the Bloch wave function, and the band dispersion, which are designated by $e^{i\mbf{k\cdot r}} |\td{u}_{n\bk}\ra$,  and $\tve_{nk} = \la \td{u}_{nk}|[\hat{H}_0 + \frac{g}{\hbar}\bs{\mu}_B\cdot \mbf{B}]|\td{u}_{nk}\ra$, respectively. Here, $n$ is the quantum number representing band index. In this Bloch basis, the interband and intraband Berry connections can be defined as $\tbc_{mn} = \la \td{u}_{m\bk}|i\pd_{\bk}|\td{u}_{n\bk}\ra$ and $\tbc_{n} = \la \td{u}_{n\bk}|i\pd_{\bk}|\td{u}_{n\bk}\ra$, respectively, with the corresponding Berry curvature given by $\td{\og}_{n\bk} = \bs{\nabla}_{\bk} \times \tbc_{n\bk}$. Moreover, the presence of a temperature gradient $\nabla T$ in the thermal Hall setup introduces an additional perturbation term $\hat{H}^{'} =-\frac{1}{2}\lbrace \hat{H}_c,\hat{r}\rbrace\cdot\et$ to the local Hamiltonian $\hat{H}_c$, where $\hat{r}$ is the position operator and $\et=-\frac{\nabla T}{T}$ represents the thermal field \cite{li2024TBCP,zhang2025TBCP}. Using time-independent perturbation theory, the first-order correction to the wave function $|\td{u}_{n\bk}\ra$ due to $\et$ field gives rise to a thermal field induced modification of the Berry curvature, $\delta^{T}\td{\og}_{n\bk} = \nabla_{\bk} \times \tbc_{n\bk}^T$ and hence the total Berry curvature upto first-order in $\et$ field is given by $(\td{\og}_{n\bk} + \delta^{T}\td{\og}_{n\bk} )$. Consequently, the Bloch state $|\td{u}_{n\bk}\ra$ also acquires a second-order energy correction $\delta^{T}\td{\ve}_{n\bk}=-\frac{1}{2}\tbc_{n}^{T}\cdot\et$. Here, $\tbc_{n}^{T}$ denotes the first-order correction to the Berry connection due to thermal field. $\tbc_{n}^{T}$ depends on the quantum geometry of the Bloch function, and is expressed as $\tbc_{n,a}^{T} = \td{\tbcp}_{ab}^{n}\et_{b}$, where $\td{\tbcp}_{ab}^{n}$ is a gauge invariant quantity known as the thermal Berry curvature polarizability (TBCP) tensor \cite{li2024TBCP,zhang2025TBCP,supp} associated with the state $|\td{u}_{n\bk}\ra$. The indices $a,\,b$ represent the cartesian components and the repeated indices are summed over. The detailed derivations of these expressions are provided in the Supplementary Material (SM) \cite{supp}. Since the correction to the Berry curvature, $\delta^{T}\td{\og}_{n\bk}$ is already linear in the thermal gradient $\nabla T$, the corresponding heat current generated by this term naturally appears at second order in $\nabla T$. The resulting nonlinear \textit{intrinsic} thermal Hall current takes the form \cite{supp,HeatCurrent2010Bergman,HeatCurrent2011Yokoyama,NATHE2020Nandy}, 


\begin{eqnarray}
    \bs{j}^{Q(\text{int})} &=& -\frac{\mathrm{k}_B^2T}{\hbar} \sum_{n} \int\lt[d\bk\rt] \lt[\bs{\nabla}T \bs{\times} \delta^T \td{\og}_{n\bk} \rt]\times \nn \\ && \Big[\frac{\lt(\tve_{n\bk}-\mu\rt)^2}{\lt(k_BT\rt)^2}  f_{0}\lt(\tve_{n\bk}\rt)  +\frac{\pi^2}{3}-\ln^2\lt(1-f_{0}\lt(\tve_{n\bk}\rt)\rt)  \nn\\ &&  - 2 \text{ Li}_2 \lt(1-f_{0}\lt(\tve_{n\bk}\rt)\rt) \Big], \label{eq:current}
\end{eqnarray}

\noindent where $f_0(\tve_{n\bk})$ is the Fermi–Dirac distribution, $\mu$ the chemical potential, $\mathrm{k}_B$ the Boltzmann constant, and $\text{ Li}_2 $ the dilogarithm function.
This expression captures the \textit{intrinsic} nonlinear planar thermal Hall current, arising purely from the thermally corrected Berry curvature and independent of scattering processes. The corresponding second-order thermal Hall response tensor is defined as $j^{Q(2,\,\text{int})}_a = \kappa_{abc}\lt(-\nabla T_b\rt) \lt(-\nabla T_c\rt) \label{eq:kappaABC}$, where $\kappa_{abc}$ encodes the nonlinear thermal conductivity components.
%
%
%
In Eq.\,(\ref{eq:current}), the material-specific information is embedded in $\td{\tbcp}_{ab}^{n}$ and the energy dispersion $\tve_{n\bk}$, both defined with respect to the magnetic-field–perturbed band structure and Bloch state $|\td{u}_{n\bk}\ra$. By straightforward expansion (see SM \cite{supp}), these quantities can be expressed perturbatively in powers of the magnetic field $B$, allowing them to be related to the original unperturbed Bloch state $|u_{n\bk}\ra$ and energy eigenvalue $\ve_{n\bk}$. 
To first order in $B$, we obtain

\begin{eqnarray}\label{eq:tdtbcp}
    \td{\tbcp}_{ab}^n &=& \tbcp_{ab}^n + \Lambda_{abc}^n B_c,
\end{eqnarray}

\noindent where $\tbcp_{ab}^n $ is the thermal Berry connection polarizability (TBCP) of the $n^{th}$ band in the basis of unperturbed Bloch state $|u_{n\bk}\ra$, and $\Lambda_{abc}^n = \partial_{B_c} \td{\tbcp}_{ab}^n\big|_{B=0}$ can be interpreted as the spin susceptibility of the TBCP, evaluated for the unperturbed Bloch state. The explicit expressions for $\tbcp_{ab}^n $ and $\Lambda_{abc}^n $ are given in Eq.\,\eqref{eq:tbcp} (we set $e=\hbar=1$), while the detailed derivations are presented in the SM \cite{supp}.

 \begin{eqnarray}\label{eq:tbcp}
     \tbcp_{ab}^{n} &=& -\text{Re}\,\sum_{m(\neq n)}\frac{\lt(\ve_{n\bk} + \ve_{m\bk}\rt)v_a^{nm} v_b^{mn} }{\lt(\ve_{n\bk} - \ve_{m\bk}\rt)^3}
 \end{eqnarray}

\begin{widetext}
    \begin{eqnarray}\label{eq:lambda}
        \Lambda_{abc}^{n} &=& \text{Re}\,\sum_{m(\neq n)}\Bigg[ \frac{ \lt( \mc{M}_c^{nn} + \mc{M}_c^{mm}\rt)v_a^{nm} v_b^{mn}}{\lt(\ve_{n\bk} - \ve_{m\bk}\rt)^3} - \frac{3 \lt(\ve_{n\bk} + \ve_{m\bk}\rt)\lt( \mc{M}_c^{nn} - \mc{M}_c^{mm}\rt)v_a^{nm} v_b^{mn}}{\lt(\ve_{n\bk} - \ve_{m\bk}\rt)^4}  \nn \\ && +  \sum_{l(\neq n)} \frac{\mc{M}_{c}^{ln} \lt( v_a^{lm}v_b^{mn} +  v_a^{nm} v_b^{ml}\rt)}{\lt(\ve_{n\bk}-\ve_{l\bk}\rt)}  \frac{\lt(\ve_{n\bk} + \ve_{m\bk}\rt) }{\lt(\ve_{n\bk} - \ve_{m\bk}\rt)^3}  +  \sum_{l(\neq m)}\frac{\mc{M}_c^{lm} \lt( v_a^{nl}v_b^{mn} + v_a^{nm} v_b^{ln}\rt)}{\ve_{m\bk}-\ve_{l\bk}} \frac{\lt(\ve_{n\bk} + \ve_{m\bk}\rt) }{\lt(\ve_{n\bk} - \ve_{m\bk}\rt)^3} \Bigg],
    \end{eqnarray}
\end{widetext}

\noindent where $v_{a}^{nm} = \la u_{n\bk} | \hat{v}_a|u_{m\bk}\ra $ denotes the velocity matrix element, $\mc{M}^{nm}=-g\mu_B \bs{s}^{nm}$ represents the spin magnetic moment matrix element, and $\bs{s}^{nm}$ is the spin operator matrix element between the unperturbed Bloch states $|u_{n\bk}\ra$ and $|u_{m\bk}\ra$. Substituting the Eqs.\,\eqref{eq:tdtbcp}-\eqref{eq:lambda} in Eq.\,\eqref{eq:current} and retaining terms up to order $\mc{O}[(\nabla T)^2 \, B]$ one obtains the \textit{intrinsic} nonlinear planar thermal Hall current as $j^{Q(2, \,\text{int})}_a = \kappa_{abcd}^{\text{int}} \lt(-\nabla_b T\rt)\,\lt(-\nabla_c T\rt)\,B_d $ and the \textit{intrinsic} NPTHE conductivity tensor is given by



\begin{eqnarray}\label{eq:kappaABCD}
    \kappa_{abcd}^{\text{int}} &=& \frac{\pi^2}{3} \frac{\mathrm{k}_B^2}{\hbar} \sum_{n} \int\lt[d\bk\rt] \lt[  \lt(\pd_b \tbcp_{ac}^{n} -\pd_a \tbcp_{bc}^{n} \rt) \mc{M}^{nn}_d  \rt. \nn \\ &&  \lt. + \lt( \Lambda_{acd}^{n} \pd_b \ve_n   -  \Lambda_{bcd}^{n} \pd_a \ve_n \rt)    \rt]  \,\delta\lt( \mu - \ve_n \rt)
\end{eqnarray}.

Here, the indices belong to cartesian components $\lbrace x, y\rbrace$ and $[d\bk]$ represents the two-dimensional Brillouin-zone integral $d^2k/(2\pi)^2$. The detailed derivations of the above expressions are provided in the SM \cite{supp}.
Evidently, the conductivity tensor $\kappa_{abcd}^{\text{int}}$ is independent of scattering time $\tau$; hence, as discussed earlier, it constitutes an \textit{intrinsic} contribution to the nonlinear thermal conductivity. This \textit{intrinsic} response is distinct from \textit{extrinsic} mechanisms and is fundamentally different in origin from the previously reported nonlinear electrical and thermal Hall conductivities \cite{NPHE2019Pan,NPHE2021Wen,PHE2021Battilomo,PHE2019Nandy}.

\noindent \textcolor{black}{\textit{Symmetry analysis and angular dependency}}--- The \textit{intrinsic} NPTHE conductivity $\kappa^{int}_{abcd}$ is rank-four, $\mc{T}$-even tensor that is antisymmetric with respect to its first two indices, i.e., $\kappa^{int}_{abcd}=-\kappa^{int}_{bacd}$ as evident from Eq.\,\eqref{eq:kappaABCD}. This antisymmetry ensures that $j_a^{Q(2,\,\text{int})}\nabla_a T = 0$, confirming $j_a^{Q(2,\,\text{int})}$ indeed represents a Hall-type current. It further implies that any component with $a=b$ must vanish and that the remaining nonzero components are related by index exchange. Consequently, in two-dimensions (2D), the tensor possesses at most four independent components, which may be conveniently chosen as $\kappa^{\text{int}}_{xyyy}$, $\kappa^{\text{int}}_{yxxx}$, $\kappa^{\text{int}}_{xyyx}$, and $\kappa^{\text{int}}_{yxxy}$ assuming a 2D sample in the $x$-$y$ plane. For concreteness, we consider a rectangular 2D sample with the indices $a$ and $b$ aligned along the crystallographic $x$ and $y$ axes. Further constraints on $\kappa^{\text{int}}_{abcd}$ arise from the crystalline point-group symmetries of the underlying two-dimensional system.

Because the NPTHE tensor is odd under spatial inversion--as the magnetic field $B$ is even under inversion, whereas both the current and the thermal-driving field (analogous to the electric field) are odd, a finite response necessarily requires broken inversion symmetry $\mc{P}$. Mirror symmetries further restrict the independent tensor components according to the orientation of the mirror plane. In particular, the presence of a horizontal mirror $\sigma_z$ forbids any \text{intrinsic} component of planar response, since such a mirror would invert the magnetic field while leaving both the thermal gradient and the heat-current directions unchanged, thereby enforcing $\kappa^{\text{int}}_{abcd}=0$. 
Consequently, a finite \textit{intrinsic} planar thermal Hall effect is possible only in point groups that simultaneously lack inversion and a horizontal mirror. There are 32 crystallographic point groups in three dimensions, of which 21 are noncentrosymmetric \cite{bilbao2006Aroyo,bradley2009mathematical}. Eighteen of these are compatible with two-dimensional (layered) symmetries relevant for planar transport, and among those 18, sixteen permit an intrinsic NPTHE -- all except $C_{3h}$ and $D_{3h}$, which are excluded by their horizontal mirror symmetry $\sigma_z$. In practice, the allowed classes include the polar point groups and the listed chiral/dihedral groups $C_n,\,C_{nv}\, (n = 1,2,3,4,6)$ and $D_n\,(n=2,3,4,6)$, as well as $S_4$ and $D_{2d}$. The effect of a symmetry operation $R$ on the \textit{intrinsic} conductivity tensor is given by the usual tensor transformation law \cite{newnham2004properties}--- $\kappa_{abcd}^{\text{int}} = \text{det}(R)\,R_{aa'}R_{bb'}R_{cc'}R_{dd'}\kappa_{a'b'c'd'}^{\text{int}}$, where $R_{aa'}$ are the matrix elements of $R$. Using this relation one can enumerate the symmetry-allowed (and forbidden) components of $\kappa^{\text{int}}_{abcd}$ for each 2D point group; these restrictions are summarized in Table~\ref{tab:symm1}.

\begin{table*}[htb!]
\caption{\label{tab:symm1} Symmetry constraint on the time reversal even $\kappa^{\text{int}}_{abcd}$ tensor from elemental and point group
symmetries pertaining to non-centrosymmetric 2D materials. ``\ding{51}'' (``\ding{55}'') symbol indicates the allowed (forbidden) tensor component corresponding to a symmetry group. The point group $C_{1v}\equiv C_s$ but with vertical mirror plane $\sigma_v$ instead of horizontal mirror $\sigma_h$.  }
\begin{ruledtabular}
\small{
\begin{tabular}{ccccccc|ccccccccc}
   &\multicolumn{6}{c|}{ Elemental crystallographic symmetries} &\multicolumn{9}{c}{ Point Groups }  \\ \hline
   & $\mc{P}$, $S_6$, $\sigma_z$ & $C_2^x, C_2^y$ & $\sigma_x, \sigma_y$ & $S_4^z$ & $C_2^z$ &$ C_{3,4,6}^z$ & $C_{1,2}$ & $C_{3,4,6}$ & $C_{1v,2v}$ & $ C_{3v,4v,6v}$ & $D_2$ & $D_{3,4,6}$ & $S_4$ & $D_{2d}$ & $C_{3h},D_{3h}$ \\ \hline
   $\kappa^{\text{int}}_{yxxx}$ & \ding{55} & \ding{55} & \ding{51} & \ding{51} & \ding{51} & \ding{51} & \ding{51} & \ding{51} & \ding{51} & \ding{51} & \ding{55} & \ding{55}  & \ding{51} & \ding{55} & \ding{55}\\
   $\kappa^{\text{int}}_{xyyy}$ & \ding{55} & \ding{55} & \ding{51} & $\kappa^{\text{int}}_{yxxx}$ & \ding{51} & $-\kappa^{\text{int}}_{yxxx}$ & \ding{51} & $-\kappa^{\text{int}}_{yxxx}$ & \ding{51} & $-\kappa^{\text{int}}_{yxxx}$ & \ding{55} & \ding{55} & $\kappa^{\text{int}}_{yxxx}$  & \ding{55} & \ding{55}\\
   $\kappa^{\text{int}}_{yxxy}$ & \ding{55} & \ding{51} & \ding{55} & \ding{51} & \ding{51} & \ding{51} & \ding{51} & \ding{51} & \ding{55} & \ding{55} & \ding{51} & \ding{51}  & \ding{51} & \ding{51} & \ding{55}\\
   $\kappa^{\text{int}}_{xyyx}$ & \ding{55} & \ding{51} & \ding{55} & $-\kappa^{\text{int}}_{yxxy}$ & \ding{51} & $\kappa^{\text{int}}_{yxxy}$ & \ding{51} & $\kappa^{\text{int}}_{yxxy}$ & \ding{55} & \ding{55}& \ding{51} & $\kappa^{\text{int}}_{yxxy}$  & $-\kappa^{\text{int}}_{yxxy}$ & $-\kappa^{\text{int}}_{yxxy}$ & \ding{55}\\
\end{tabular}
}
\end{ruledtabular}
\end{table*}

\begin{figure*}[htb!]
\centering
\includegraphics[width=\linewidth]{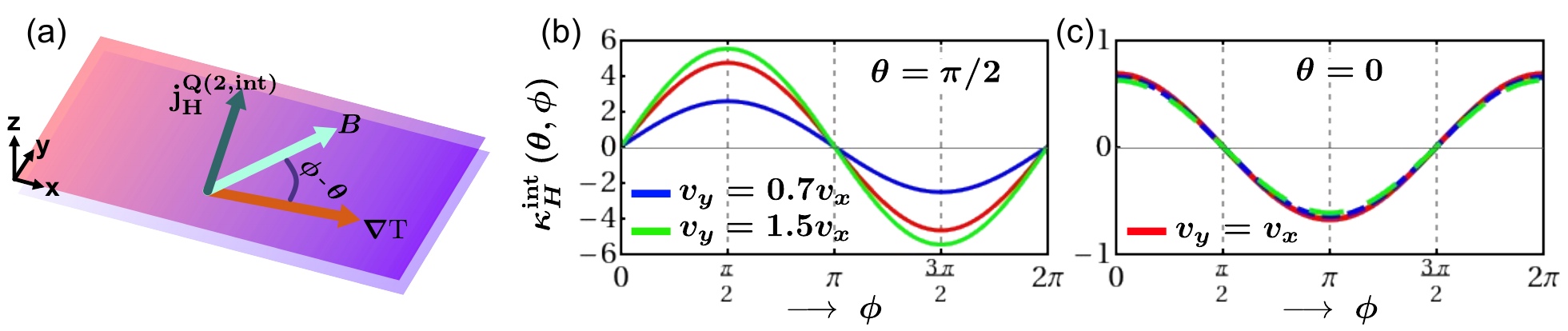}
\caption{(a) Schematic illustration of the planar thermal Hall geometry, where the temperature gradient $\bs{\nabla}T$ and magnetic field $\mbf{B}$ lie in the same plane, forming polar angles $\theta$ and $\phi$ with respect to the $x$-axis.
(b) Angular dependence of the intrinsic nonlinear planar thermal Hall coefficient $\kappa_{H}^{\text{int}}(\theta=\pi/2,\phi)$, shown in units of $ \frac{\mathrm{k}_B^2\mu_B}{\hbar}\mathrm{\AA \,eV^{-1}}$, calculated for the tilted 2D Dirac model with different anisotropy ratios: $v_y=v_x$ (red), $v_y=0.7v_x$ (blue), and $v_y=1.5v_x$ (green). (c) Same as (b), but for $\theta=0$. The parameters used are $v_x = 1\times10^6~\mathrm{m/s} \equiv 6.582~\mathrm{eV\AA}$, $m=0.02~\mathrm{eV}$, $t=0.1v_x$, and $\mu=0.05~\mathrm{eV}$.}
\label{fig:HallBar}
\end{figure*}

In a prototype experimental setup with coplanar $\bs{\nabla} T$ and $\mbf{B}$ field as shown in Fig.\,\ref{fig:HallBar}(a), we assume $\bs{\nabla} T=\lt(\nabla_xT\cos\theta, \nabla_yT\sin\theta\rt)$ and $\mbf{B}=\lt(B_x\cos\phi,B_y\sin\phi\rt)$ makes polar angle $\theta$ and $\phi$ respectively, with the $x$-axis. In this configuration, the polar-angle-dependent \textit{intrinsic} NPTHE current flows along $\hat{z}\times \bs{\nabla} T$ and is given by

\begin{eqnarray}
    j^{Q(2, \,\text{int})}_H \lt(\theta,\phi\rt) &=& \kappa_{H}^{\text{int}}  \lt(\theta,\phi\rt)\,\lt(\nabla T\rt)^2\,B \,,
\end{eqnarray}

where the angle-dependent coefficient $\kappa_{H}^{\text{int}}  \lt(\theta,\phi\rt)$ is expressed in terms of the components $\kappa_{abcd}^{\text{int}}$ as

\begin{eqnarray} \label{eq:kappaAngular}
    \kappa_{H}^{\text{int}}  \lt(\theta,\phi\rt) &=& \lt(\kappa_{yxxx}^{\text{int}}\cos\phi + \kappa_{yxxy}^{\text{int}}\sin\phi\rt)\cos\theta\, \nn \\
    && - \lt(\kappa_{xyyx}^{\text{int}}\cos\phi + \kappa_{xyyy}^{\text{int}}\sin\phi \rt)\sin\theta \,.
\end{eqnarray}

\noindent This formulation naturally captures the dependence of the \textit{intrinsic} NPTHE current on the relative orientation of the thermal gradient and magnetic field, which is determined by the angle difference $\phi-\theta$. As summarized in Table~\ref{tab:symm1}, it is evident that for $n \geq 3$, only one independent tensor component remains nonzero in the polar point groups $C_{nv}$ and the dihedral groups $D_n$. Accordingly, the angular dependence of the \textit{intrinsic} conductivity simplifies to: $\kappa_{H}^{\mathrm{int}}(\theta,\phi)
= \kappa_{yxxx}^{\mathrm{int}} \cos(\theta - \phi)
 \text{ for } C_{nv}\ (n \ge 3)$, and  $\kappa_{H}^{\mathrm{int}}(\theta,\phi)
= \kappa_{yxxy}^{\mathrm{int}} \sin(\phi - \theta)
 \text{ for } D_n\ (n \ge 3)$. These expressions highlight the characteristic cosine and sine angular dependences that distinguish the symmetry behavior of the \textit{intrinsic} NPTHE in polar and dihedral point groups, respectively.




\noindent \textcolor{black}{\textit{Intrinsic nonlinear transport coefficients in Dirac system}}---  Considering a generic tilted 2D Dirac model Hamiltonian in Eq.\,\eqref{eq:Ham}, we have conducted an analytical and numerical assessment of $\kappa_{H}^{\mathrm{int}}(\theta,\phi)$.

\begin{eqnarray} \label{eq:Ham}
    H_s &=& t_s k_x \sigma_0 + v_y k_y \sigma_x + s v_x k_x \sigma_y + m\sigma_z ,
\end{eqnarray} 

where $s=\pm$ is the valley index, $t_s = st$ parametrizes the valley dependent tilt, $v_{x,y}$ are the Fermi velocities along the principal axes, $m$ is the Dirac mass that opens an energy gap, $\sigma_i$'s are the Pauli matrices, and $\sigma_0$ is the identity matrix. This minimal model captures the low-energy physics of a broad class of two-dimensional Dirac materials \cite{RevModPhys2010ZahidHasan,RevModPhys2011Xiao}, including the surfaces of topological crystalline insulators \cite{TCI2013LiangFu,TCI2014LiangFu,TCI2013Okada,TCI2012Hsieh} and strained transition-metal dichalcogenides \cite{WTe2018Du,TMDC2019Zhou,TMDC2020Zhou}. The inclusion of the tilt parameter $t$ in Eq.\,\eqref{eq:Ham} breaks both rotational and inversion symmetries. When both valleys are considered together, the full Hamiltonian retains a mirror symmetry $M_x : (k_x \rightarrow -k_x)$, which interchanges the two valleys. For vanishing mass term each individual valley separately preserves an additional mirror symmetry $M_y$. These mirror symmetries play crucial roles in determining the allowed components of the \textit{intrinsic} NPTHE as discussed in Table\,\ref{tab:symm1}. The energy dispersion of the Hamiltonian is given by $\ve_{k}^{\pm} = stk_x \pm (m^2 +k_x^2 v_x^2 + k_y^2 v_y^2),^{1/2}$ where $\pm$ denotes the conduction and valence band, respectively. The TBCP tensor for this Dirac Hamiltonian can be directly evaluated using Eq.\,\eqref{eq:tbcp}. Interestingly, the TBCP is proportional to the valley-dependent tilt parameter $t_s$, in contrast to the electric field induced BCP~\cite{BCP2022Liu,IPHE2023,kamaldas2024valleyHall}. The explicit analytical expressions for the components of the TBCP tensors are provided in the SM \cite{supp}. Note that, owing to the mirror symmetry $M_x$, which interchanges the two valleys, the \textit{intrinsic} NPTHE tensor satisfies $ \kappa_{xyyy}^{\text{int}} (s,k_x,k_y) = - \kappa_{xyyy}^{\text{int}} (-s,-k_x,k_y) $ as can be directly verified from the analytical structure of the integrand in Eq.\,\eqref{eq:kappaABCD}. This relation implies that the valley-resolved contributions of $\kappa_{xyyy}^{\text{int}}$ are opposite in sign, i.e., $ \kappa_{xyyy}^{\text{int}} (s) = - \kappa_{xyyy}^{\text{int}} (-s)$ due to the presence of $M_x$ mirror. In contrast, the component $\kappa_{yxxx}^{\text{int}}$ remains even under valley exchange, satisfying $\kappa_{yxxx}^{\text{int}} (s) =  \kappa_{yxxx}^{\text{int}} (-s)$. Furthermore, the integrand of $\kappa_{abba}^{\text{int}} (s,k_x,k_y)$ is odd in $k_y$, and therefore its contribution vanishes upon Brillouin-zone integration. Focusing on the $s=+$ valley, the analytical expressions for the components of $\kappa_{abbb}^{\text{int}}$ in the conduction band (chemical potential $\mu>0$) for the isotropic case $v_x=v_y=v$ are obtained as

\begin{eqnarray*}
    \kappa_{yxxx}^{\text{int}} &=& \frac{g \pi \mu_B \mathrm{k}_B^2\lt( 7m^4 -10m^2\mu^2 + 3 \mu^4 \rt) t^2 }{384\hbar v \mu^6} \\
    \kappa_{xyyy}^{\text{int}} &=& -\frac{g\pi \mu_B \mathrm{k}_B^2\lt(3m^4 -26m^2\mu^2 + 15 \mu^4\rt) t^2}{384\hbar v \mu^6}
\end{eqnarray*}

\noindent These analytical expressions reveal that both the \textit{intrinsic} components scale quadratically with the tilt parameter $t$, and vanishes in the untilted limit. Moreover, the strong dependence on the ratio $m/\mu$ indicates that the response is most pronounced near the band edge, where band geometric effects are enhanced. In the limit $|\mu|\gg m$, both components decay rapidly as $\mu^{-2}$, signifying the suppression of quantum geometric effects deep inside the conduction band. Thus, the magnitude of the nonlinear planar thermal Hall response can be effectively tuned by the tilt strength, chemical potential, and band gap. To further substantiate these analytical findings, we have numerically evaluated $\kappa_{abcd}^{\text{int}}$ for the anisotropic case ($v_x \ne v_y$), which exhibits similar dependencies on $t$ and $\mu$ as in the isotropic limit ($v_x = v_y$). The angular variation of the \textit{intrinsic} NPTHE $\kappa_{H}^{\mathrm{int}}(\theta,\phi)$ for different anisotropy ratios $v_y/v_x$, shown in Fig.~\ref{fig:HallBar}(b,c), displays a characteristic modulation governed by the relative orientation between the thermal gradient and the magnetic field for both isotropic and anisotropic cases. At oblique orientations ($\theta= \pi/2$) of the temperature gradient with respect to the $x$-axis, the anisotropy in Fermi velocity produces distinct amplitude variations in  $\kappa_{H}^{\mathrm{int}}(\theta,\phi)$ reflecting the reduced rotational symmetry of the tilted Dirac dispersion, as shown in Fig.~\ref{fig:HallBar}(b). In contrast, for $\theta= 0$, where the thermal gradient is aligned along the $x$-axis, all curves--red ($v_y=v_x$), blue ($v_y=0.7v_x$), and green ($v_y=1.5v_x$)--collapse onto each other, indicating that the \textit{intrinsic} NPTHE becomes insensitive to in-plane Fermi-velocity anisotropy when Hall transport occurs along $y$-axis, as illustrated in Fig.~\ref{fig:HallBar}(c).

\noindent \textit{Outlook}--- The \textit{intrinsic} NPTHE proposed here can be experimentally distinguished from other nonlinear Hall responses--such as the Berry curvature dipole driven nonlinear Hall effect \cite{NATHE2020Nandy,NTHE2022Nandy} or \textit{extrinsic} skew-scattering \cite{NTHE2022Zhou,barman2025thirdorder} contributions--through its distinct scaling behavior, symmetry dependence, and dissipationless nature. First, unlike \textit{extrinsic} mechanisms that depend on the relaxation time $\tau$ \cite{NATHE2020Nandy,NTHE2022Nandy,NTHE2022Zhou,INTHE2023Harsh,LNPHE2021Wen,NPHE2019Pan}, the \textit{intrinsic} NPTHE originates purely from the quantum geometric correction to the Berry curvature and is therefore independent of $\tau$, similar to the \textit{intrinsic} anomalous Hall and thermal Hall effects \cite{Nagaosa2010Anomalous, Xiao2010Berry}. This distinction can be verified experimentally by examining its temperature and disorder dependence: a $\tau$-independent response persisting in the clean limit signals intrinsic origin. Second, the planar configuration--with the magnetic field and temperature gradient lying in the same plane--provides a clear diagnostic distinction from ordinary thermal Hall effects, where the field is perpendicular to the transport plane. The angular modulation of the Hall voltage with respect to the relative orientation of the thermal and magnetic fields, following the $\cos(\phi-\theta)$ or $\sin(\phi-\theta)$ dependence predicted in Eq.~\eqref{eq:kappaAngular}, would serve as a direct signature of the planar geometry. Furthermore, the symmetry selection rules derived earlier imply that the signal should vanish in centrosymmetric or mirror-symmetric crystals (e.g., $C_{3h}$, $D_{3h}$), but appear exclusively in polar or chiral point groups lacking horizontal mirror symmetry--providing an additional symmetry-based criterion for identification. Together, these features offer experimentally accessible means to isolate and confirm the \textit{intrinsic} NPTHE as a distinct geometric nonlinear transport phenomenon.

\noindent \textit{Conclusion}--- We have developed the theory of the \textit{intrinsic} NPTHE, a second-order in temperature gradient, dissipationless thermal response arising from a thermal gradient induced correction to the Berry curvature, characterized by the thermal Berry connection polarizability tensor.
Our symmetry analysis establishes that the effect is allowed only in noncentrosymmetric systems lacking horizontal mirror symmetry, and our model study based on a tilted Dirac Hamiltonian demonstrates its tunability via tilt strength, chemical potential, and anisotropy. The $\tau^0$ scaling and characteristic angular dependence underscore its intrinsic quantum geometric nature, setting it apart from extrinsic nonlinear Hall responses.

{\par} I am grateful to Dr. Snehasish Nandy, Dr. Surajit Sarkar, Dr. Hridis K Pal, and Bishal Das for valuable discussions. I acknowledge the Department of Physics at
University of Cagliari, Italy and Dr. Fabio Bernardini, \& Dr. Alessio Filippetti for providing the necessary facilities. The author acknowledges support from the Italian Ministry of University and Research (MUR), financed by the European Union – Next Generation EU, through the PRIN 2022 project SUBLI ``Sustainable spin generators based on Van der Waals dichalcogenides,'' contract No. 2022M3WXE7. Additional support from the PRIN 2022 TOTEM project, grant No. F53D23001080006, funded by MUR, is also acknowledged.

\bibliography{ref}
\end{document}